\documentclass[letterpaper]{article} % DO NOT CHANGE THIS
\usepackage{aaai23}  % [submission] DO NOT CHANGE THIS
\usepackage{times}  % DO NOT CHANGE THIS
\usepackage{helvet}  % DO NOT CHANGE THIS
\usepackage{courier}  % DO NOT CHANGE THIS
\usepackage[hyphens]{url}  % DO NOT CHANGE THIS
\usepackage{graphicx} % DO NOT CHANGE THIS
\usepackage{natbib}  % DO NOT CHANGE THIS AND DO NOT ADD ANY OPTIONS TO IT
\usepackage{caption} % DO NOT CHANGE THIS AND DO NOT ADD ANY OPTIONS TO IT
\usepackage{algorithm}
\usepackage{algorithmic}
\usepackage{amsmath}
\usepackage[usestackEOL]{stackengine}

\usepackage{newfloat}
\usepackage{listings}
\DeclareCaptionStyle{ruled}{labelfont=normalfont,labelsep=colon,strut=off} % DO NOT CHANGE THIS
\floatstyle{ruled}
\newfloat{listing}{tb}{lst}{}
\floatname{listing}{Listing}
\title{Multifidelity Active Learning for Failure Estimation of TRISO Nuclear Fuel}
\author{
    Somayajulu L. N. Dhulipala, \textsuperscript{\rm 1}
    Promit Chakroborty, \textsuperscript{\rm 2}
    Michael D. Shields, \textsuperscript{\rm 2} \\
    Wen Jiang, \textsuperscript{\rm 1}
    Benjamin W. Spencer, \textsuperscript{\rm 1}
    and Jason D. Hales \textsuperscript{\rm 1}
}
\affiliations{
    \textsuperscript{\rm 1}
    Idaho National Laboratory\\

    \textsuperscript{\rm 2}
    Johns Hopkins University\\
    
}

\usepackage{bibentry}
\begin{document}

\maketitle

\begin{abstract}
The Tristructural isotropic (TRISO)-coated particle fuel is a robust nuclear fuel proposed to be used for multiple modern nuclear technologies. Therefore, characterizing its safety is vital for the reliable operation of nuclear technologies. However, the TRISO fuel failure probabilities are small and the computational model is time consuming to evaluate them using traditional Monte Carlo-type approaches. In the paper, we present a multifidelity active learning approach to efficiently estimate small failure probabilities given an expensive computational model. Active learning suggests the next best training set for optimal subsequent predictive performance and multifidelity modeling uses cheaper low-fidelity models to approximate the high-fidelity model output. After presenting the multifidelity active learning approach, we apply it to efficiently predict TRISO failure probability and make comparisons to the reference results.
\end{abstract}

\section{Introduction}

The Tristructural isotropic (TRISO)-coated particle fuel is an advanced nuclear fuel designed to withstand extreme operating temperatures inside reactors \cite{Williamson2021a, Jiang2021a}. Due to its robustness, it is being proposed to be used for multiple modern nuclear technologies, such as small modular reactors, advanced reactors, and microreactors.
Each TRISO particle has a fuel kernel and several protective outer layers, and a particle is considered to have failed if the outer protective layer has fractured due to thermo-mechanical stresses \cite{Jiang2021a}. Due to the uncertainties in the fuel properties and the particle geometries, the failure of a TRISO particle is characterized probabilistically, and the failure probabilities can range from $10^{-3}$ to $10^{-6}$ \cite{Dhulipala_TRISO}.

% \cite{Dhulipala_AL_TRISO}

Failure probability estimation involves estimating the following integral equation or its discretized form which is also presented below:

\begin{equation}
    \label{eqn:Pf_exact_approx}
    P_f = \int_{{F}(\pmb{X}) > \mathcal{F}} q(\pmb{X})~d\pmb{X} \approx \frac{1}{N_m}~\sum \mathbf{I}\big({F}(\pmb{X}) > \mathcal{F}\big)
\end{equation}

\noindent where $\pmb{X}$ are the uncertain inputs, $q(.)$ defines their probability distributions, $F(\pmb{X})$ is the model value, and $\mathcal{F}$ is the failure threshold. In the discretized form, $I(.)$ is an indicator function when model value exceeds the failure threshold and $N_m$ are the number of model evaluations. The discretization in Equation \eqref{eqn:Pf_exact_approx} represents a Monte Carlo approach to compute the failure probability and this approach is computationally infeasible when the failure probabilities are small and the model $F(.)$ is time consuming to evaluate (which is the case for TRISO fuel model). Although variance reduction methods like importance sampling and subset simulation can reduce the number of model evaluations, they are still computationally expensive when the model itself is time consuming. Therefore, in this paper, we present an active learning approach coupled with multifidelity modeling to accelerate the failure probability estimation (adapted from \citet{Dhulipala_AL_MFM, Dhulipala_AL_TRISO}). Active learning is a subset of artificial intelligence where the machine learning model actively suggests the next best training set that would lead an optimal subsequent predictive performance. Multifidelity modeling uses multiple low-fidelity (LF) models that are cheaper to evaluate to approximate the high-fidelity (HF) model output. We first present the active learning with multifidelity modeling approach and then apply it to efficiently and accurately estimate the TRISO fuel failure probability.

\section{Methods}\label{sec:methods}

\subsection{Subset simulation}

Subset simulation expresses the failure probability $\hat{P}_f$ as a product of larger failure probabilities:

\begin{equation}
    \label{eqn:SS1}
    \begin{aligned}
        P_f &= P({F}(\pmb{X})>\mathcal{F}_1) \prod_{s=2}^{N_s}~P({F}(\pmb{X})>\mathcal{F}_s|{F}(\pmb{X})>\mathcal{F}_{s-1}) \\
        &\equiv P_1 \prod_{s=2}^{N_s}~P_{s|s-1} \\
    \end{aligned}
\end{equation}

\noindent where ${P}_1$ is the first unconditional failure probability computed as the fraction of samples exceeding the intermediate failure threshold, $\mathcal{F}_1$, and ${P}_{s|s-1}$ are the subsequent conditional failure probabilities that are conditional on exceeding the prior intermediate failure thresholds, $\mathcal{F}_{s-1}$, and are computed as the fraction of samples exceeding the intermediate failure threshold $\mathcal{F}_s$. In expressing ${P}_f$ as a product of larger failure probabilities, subset simulation creates intermediate failure thresholds $(\mathcal{F}_1, \dots, \mathcal{F}_{N_s-1})$ before the required threshold of $\mathcal{F}$. The samples falling in between two subsequent intermediate failure thresholds $\mathcal{F}_{s-1}$ and $\mathcal{F}_{s}$ constitute a subset. While a regular Monte Carlo simulation (MCS) is used to estimate ${P}_1$, a Markov chain Monte Carlo (MCMC) is used to estimate ${P}_{s|s-1}$. To estimate these intermediate failure probabilities, the intermediate failure thresholds must be specified. These are usually taken as the $1-p_o$ percentile value of sample outputs from each subset. $p_o$ is fixed \emph{a priori} (0.1 is usually used). A component-wise Metropolis-Hastings algorithm (an MCMC algorithm) proposed by \citet{Au2001a} is popular for simulating the conditional samples in ${P}_{s|s-1}$.

\subsection{Gaussian process active learning within subset simulation}\label{sec:al_ss}

A Gaussian process (GP) is a Bayesian surrogate model that models the conditional probability distribution of the form \cite{Rasmussen2004}:

\begin{equation}
    \label{eqn:Kriging_1}
    \begin{aligned}
    &p(\pmb{y}_*~|~\pmb{X}, \pmb{X}_*, \pmb{y}) \sim \\ 
    &\mathcal{N}\Big(~k(\pmb{X}_*,\pmb{X})~k(\pmb{X},\pmb{X})^{-1}~\pmb{y}, \\
    & k(\pmb{X}_*,\pmb{X}_*) - k(\pmb{X}_*,\pmb{X})~k(\pmb{X},\pmb{X})^{-1}~k(\pmb{X},\pmb{X}_*)~\Big) \\
    \end{aligned}
\end{equation}

\noindent where, $(\pmb{X}, \pmb{y})$ is the training data, $\pmb{X}_*$ is the test inputs, and $p(\pmb{y}_*~|~.)$ is the distribution of the test predictions modeled by the GP. $k(.,~.)$ in the above equation is a kernel covariance function usually defined by a squared exponential function \cite{Dhulipala_AL_TRISO}. $p(\pmb{y}_*~|~.)$ modeled by the GP is follows a Normal distribution with mean $\mu_{{\mathcal{G}}} = k(\pmb{X}_*,\pmb{X})~k(\pmb{X},\pmb{X})^{-1}~\pmb{y}$ and variance $\sigma_{{\mathcal{G}}}^2 = k(\pmb{X}_*,\pmb{X}_*) - k(\pmb{X}_*,\pmb{X})~k(\pmb{X},\pmb{X})^{-1}~k(\pmb{X},\pmb{X}_*)$. The kernel function has hyper-parameters that need to be optimized and this optimization is performed by maximizing the negative log-likelihood of observing the training data.  

The computational model evaluations required for subset simulation in Equation \eqref{eqn:SS1} can be replaced by a GP mean prediction. In addition, since a GP provides uncertainty estimates in the form of $\sigma_{{\mathcal{G}}}$ in Equation \eqref{eqn:Kriging_1}, it can be used to device an active learning scheme. The scheme reverts to calling the expensive computational model whenever the GP uncertainty estimate is large given test inputs $\pmb{X}_*$. Specifically, in subset simulation we are interested in characterizing the intermediate and final failure thresholds (i.e., $\mathcal{F}_s$ and $\mathcal{F}$). Therefore, we can design an active learning function which becomes sensitive near the intermediate and final failure thresholds. \citet{Dhulipala_AL_MFM} propose one such active learning function based an the U-learning function defined as:

\begin{equation}
    \label{eqn:U_func1}
    U_s = \frac{|\mu_{{\mathcal{G}}}-\mathcal{F}_s|}{\sigma_{{\mathcal{G}}}}~~\forall~s<N_s
\end{equation}

\noindent Whenever the GP predictions are close to an intermediate or the final failure threshold or the GP standard deviation is large, the U-function value will be small. A U-function value less than 2.0 \cite{Echard2011a} is when the computational model is called and the GP is re-trained. U-function value of 2.0 has about $2.5\%$ probability of the GP mis-characterizing failure as safe and vice-versa \cite{Echard2011a}. By only calling the expensive computational model when needed (typically near the intermediate and final failure thresholds) and approximating the model evaluations with GP predictions elsewhere, active learning can bring significant computational benefits for estimating small failure probabilities.

\subsection{Use of multiple modeling fidelities}

Instead of relying only on the HF model in the active learning procedure described in Section \ref{sec:al_ss}, we can create a suite of LF models to the HF model and perform active learning with the LF models and the HF model for even more computational gains. In the multifidelity modeling setting, given a suite of $M$ LF models $f_i(\pmb{X})~(i \in \{1, \dots, M\})$ and the HF model $F(\pmb{X})$, a GP is trained to learn the difference between each of the LF and HF model:

\begin{equation}
    \label{eqn:MF1}
    \varepsilon^i = F(\pmb{X}) - f_i(\pmb{X})
\end{equation}

\noindent where, $\varepsilon^i$ is the difference between the HF model and LF model $i$. Once trained, given a test input $\pmb{X}_*$, the GP for LF model $i$ predicts distribution of the correction term with mean and standard deviation:

\begin{equation}
    \label{eqn:MF2}
    p(\varepsilon^i_*~|~\pmb{X}, \pmb{X}_*, \varepsilon^i) \sim \mathcal{N}(\mu^i_G,~(\sigma^i_G)^2)
\end{equation}

\noindent Given the predictive distributions of the correction terms for $M$ LF models (i.e., $\varepsilon^i_*$), we select that LF model which has the greatest probability of having the least magnitude of the correction term. Since the GP correction terms are distributions rather than scalar values, this selection scheme is mathematically represented as:

\begin{equation}
    \label{eqn:MF3}
    w_i = P(|\varepsilon^i_*| = \textrm{min}_i|\varepsilon^i_*|)
\end{equation}

\noindent where, $w_i$ is the weight given to the LF model $i$ based upon its GP correction magnitude being the smallest among the $M$ LF models and $P(.)$ is the probability. Since $\varepsilon^i_*$ follows a Normal distribution and $|\varepsilon^i_*|$ follows a folded Normal distribution, a mathematical expression can be derived for Equation \eqref{eqn:MF3} as: 

\begin{equation}
    \label{eqn:MF4}
    w_i = \int_0^\infty \bigg[p_{|\varepsilon^i_*|}(z) \prod_{k \neq i} \big\{1 - P_{|\varepsilon^k_*|}(z)\big\}\bigg] dz
\end{equation}

\noindent where, $p_{|\varepsilon^i_*|}(z)$ is the probability density of a folded Normal distribution and $P_{|\varepsilon^i_*|}(z)$ is its cumulative distribution function. The LF model with the highest $w_i$ value can be selected and evaluated. The corrected LF model value using the mean correction term is then expressed as:

\begin{equation}
    \label{eqn:MF5}
    F(\pmb{X}_*) \approx f_i(\pmb{X}_*) + \mu^i_G
\end{equation}

\noindent In addition, the U-function that aids in deciding whether or not to call the HF model can be modified from Equation \eqref{eqn:U_func1} as:

\begin{equation}
    \label{eqn:MF6}
    U_s = \frac{|f_i(\pmb{X}_*) + \mu^i_G-\mathcal{F}_s|}{\sigma^i_G}~~\forall~s<N_s
\end{equation}

\noindent where the LF model $i$ prediction and the associated GP correction have been used in the numerator.

Given $M$ LF models with varying computational complexities $\tau_i~(i \in \{1, \dots, M\})$, it also possible to incorporate the LF model cost into the selection scheme described in Equations \eqref{eqn:MF1}-\eqref{eqn:MF4}. Consider a cost function of the form:

\begin{equation}
    \label{eqn:MF7}
    \gamma(\tau_i) = \tau_i^\beta > 0
\end{equation}

\noindent where, $\beta > 0$ is a tuning parameter. Now, the selection scheme using the weights in Equation \eqref{eqn:MF3} can be be modified to consider the model cost as follows:

\begin{equation}
    \label{eqn:MF8}
    w_i = P(|\gamma(\tau_i)~\varepsilon^i_*| = \textrm{min}_i|\gamma(\tau_i)~\varepsilon^i_*|)
\end{equation}

\noindent Equation \eqref{eqn:MF8} still follows the expression in Equation \eqref{eqn:MF4} but with $|\varepsilon^i_*|$ multiplied by the factor $\gamma(\tau_i)$.

\section{Multifidelity TRISO fuel modeling}

\begin{figure}[h]
\centering
\includegraphics[width=0.9\columnwidth]{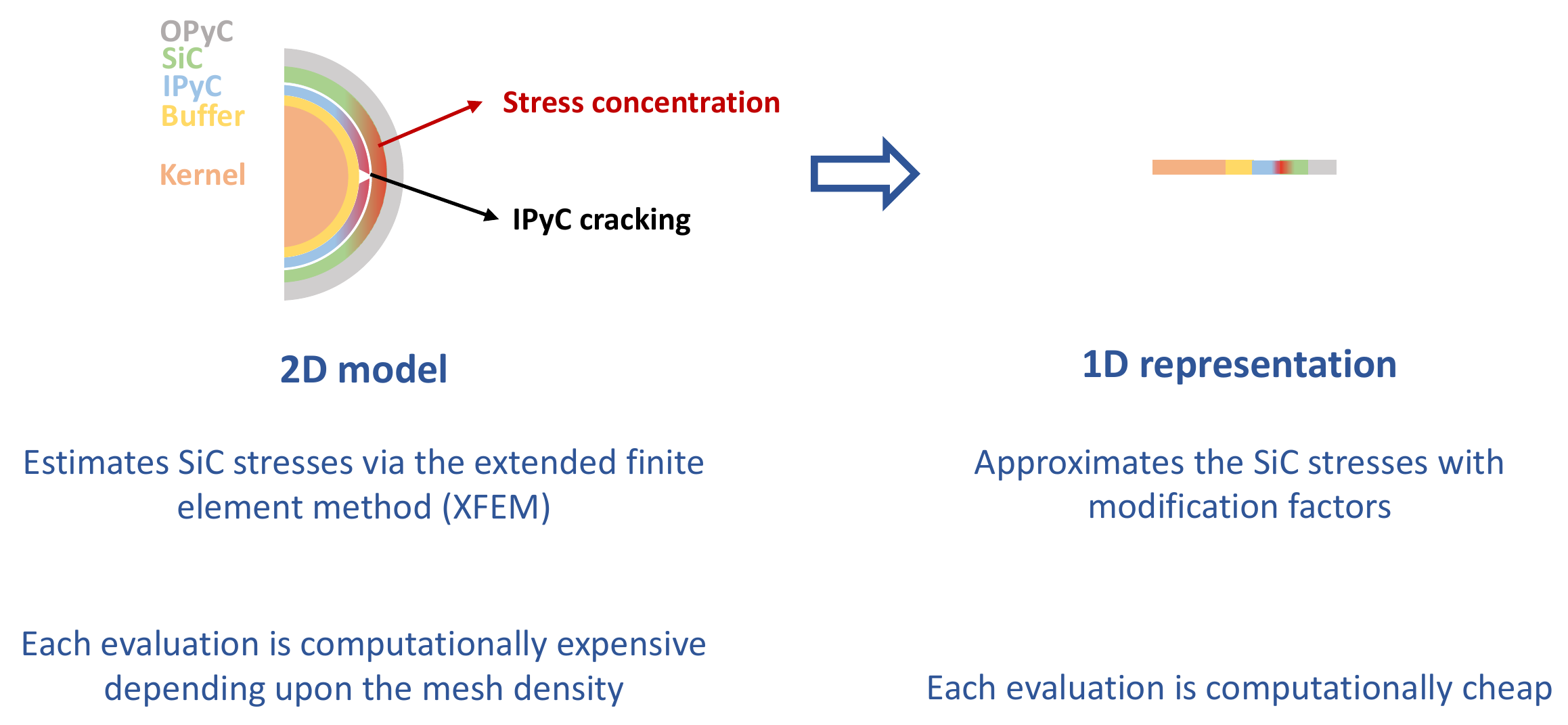} % Reduce the figure size so that it is slightly narrower than the column. Don't use precise values for figure width.This setup will avoid overfull boxes.
\caption{2-D and 1-D representations of the TRISO particle fuel with a description of the stress calculation approach. This figure is adapted from \citet{Dhulipala_TRISO}.}
\label{fig:triso_models}
\end{figure}

The TRISO particle is composed of a fuel kernel and four layers: buffer, inner pyrolytic carbon (IPyC), silicon carbide (SiC), and outer pyrolytic carbon (OPyC), as presented in Figure \ref{fig:triso_models}. Due to fission and the generated heat in the fuel kernel, thermo-mechanical stresses are exerted on the outer layers of the TRISO particle. When the stresses exceed the strength, the IPyC layer cracks leading to a stress concentration in the SiC layer which can further lead to the failure of the SiC layer. The TRISO particle is considered to have failed when the SiC layer fails.

% , which is built using the Multiphysics Object Oriented Simulation Environment \cite{Permann2020a},

To model the failure of the TRISO particle (via IPyC and SiC layers cracking), multiple fidelities are possible, as presented in Figure \ref{fig:triso_models}. We use the Bison fuel performance code \cite{Williamson2021a} to model the TRISO particle failure. The 2-D modeling approach explicitly simulates a crack in the IPyC layer using the extended finite element method (XFEM) and estimates the stresses in the SiC layer caused due to stress concentration \cite{Jiang2021a}. Such an approach can accurately model the TRISO particle behavior but it is also computationally expensive depending upon the mesh density of the model. Alternatively, the 1-D modeling approach is computationally cheap. Since IPyC cracking cannot be modeled in 1-D, stresses in the IPyC and SiC layers are approximated via stress modification factors. Therefore, although the 1-D approach is computationally efficient, it only provides an approximation of the TRISO particle failure.

Several uncertainties exist in modeling the TRISO particle. These are related to the geometrical uncertainties of the layer thicknesses (i.e., buffer, IPyC, SiC, and OPyC) and the strength uncertainties of the IPyC and SiC layers. Due to these uncertainties, failure of the SiC layer is characterized probabilistically. In general, the failure probabilties are small and a direct application of the Monte Carlo method can be computationally infeasible \cite{Dhulipala_TRISO}.

\section{Results}

The failure probability of TRISO fuel is estimated using the multifidelity active learning procedure presented in Section \ref{sec:methods} and compared to the reference result. Two cases of multiple modeling fidelities are considered: (1) one LF and one HF model; and (2) two LF models and one HF model. In the first case, the LF model is the 1-D TRISO model and the HF model is the 2-D TRISO model with a coarse mesh. In the second case, the two LF models are 1-D (LF1) and 2-D coarse mesh (LF2) and the HF model is the 2-D fine mesh. The results corresponding to these cases are presented below. 

\subsection{Case 1: One LF (1-D) model and one HF (2-D coarse mesh) model}

Table \ref{table2} presents the failure probability results of the TRISO particle. It is noticed that the failure probability estimated by the multifidelity active learning method matches with that of the reference method for a similar coefficient of variation (COV) value. In addition, the number of HF (2-D coarse mesh) TRISO model evaluations are substantially smaller for multifidelity active learning than for the reference. 

\begin{table}[h]
\centering
\caption{Failure probability and coefficient of variation (COV) estimated by the multifidelity active learning and reference methods. The corresponding number of HF model evaluations (i.e., 2-D coarse mesh) are also reported. These results are adapted from \citet{Dhulipala_AL_TRISO}.}
\begin{tabular}{|c|c|c|c|}
\hline
     & \textbf{Failure prob.} & \textbf{COV} & \textbf{$\#$ model evals.} \\
    \hline
    \Centerstack[c]{\textbf{MF active} \\ \textbf{learning}} & $6.76E-4$ & $0.073$ & $190$ \\
    \hline
    \Centerstack[c]{\textbf{Reference}} & $6.78E-4$ & $0.071$ & $20,000$ \\
    \hline
\end{tabular}
\label{table2}
\end{table}

\subsection{Case 2: Two LF (1-D and 2-D coarse mesh) models and one HF (2-D fine mesh) model}

Table \ref{table3} presents the failure probability results of the TRISO particle. It is noticed that the failure probability estimated by the multifidelity active learning method matches with that of the reference method for a similar coefficient of variation (COV) value. In addition, the number of HF (2-D fine mesh) TRISO model evaluations are again substantially smaller for multifidelity active learning than for the reference.

\begin{table}[h]
\centering
\caption{Failure probability and coefficient of variation (COV) estimated by the multifidelity active learning and reference methods. The corresponding number of HF model evaluations (i.e., 2-D fine mesh) are also reported.}
\begin{tabular}{|c|c|c|c|}
\hline
     & \textbf{Failure prob.} & \textbf{COV} & \textbf{$\#$ model evals.} \\
    \hline
    \Centerstack[c]{\textbf{MF active} \\ \textbf{learning}} & $1.17E-2$ & $0.09$ & $58$ \\
    \hline
    \Centerstack[c]{\textbf{Reference}} & $0.99E-2$ & $0.07$ & $10,000$ \\
    \hline
\end{tabular}
\label{table3}
\end{table}

\begin{figure}[h]
\centering
\includegraphics[width=0.9\columnwidth]{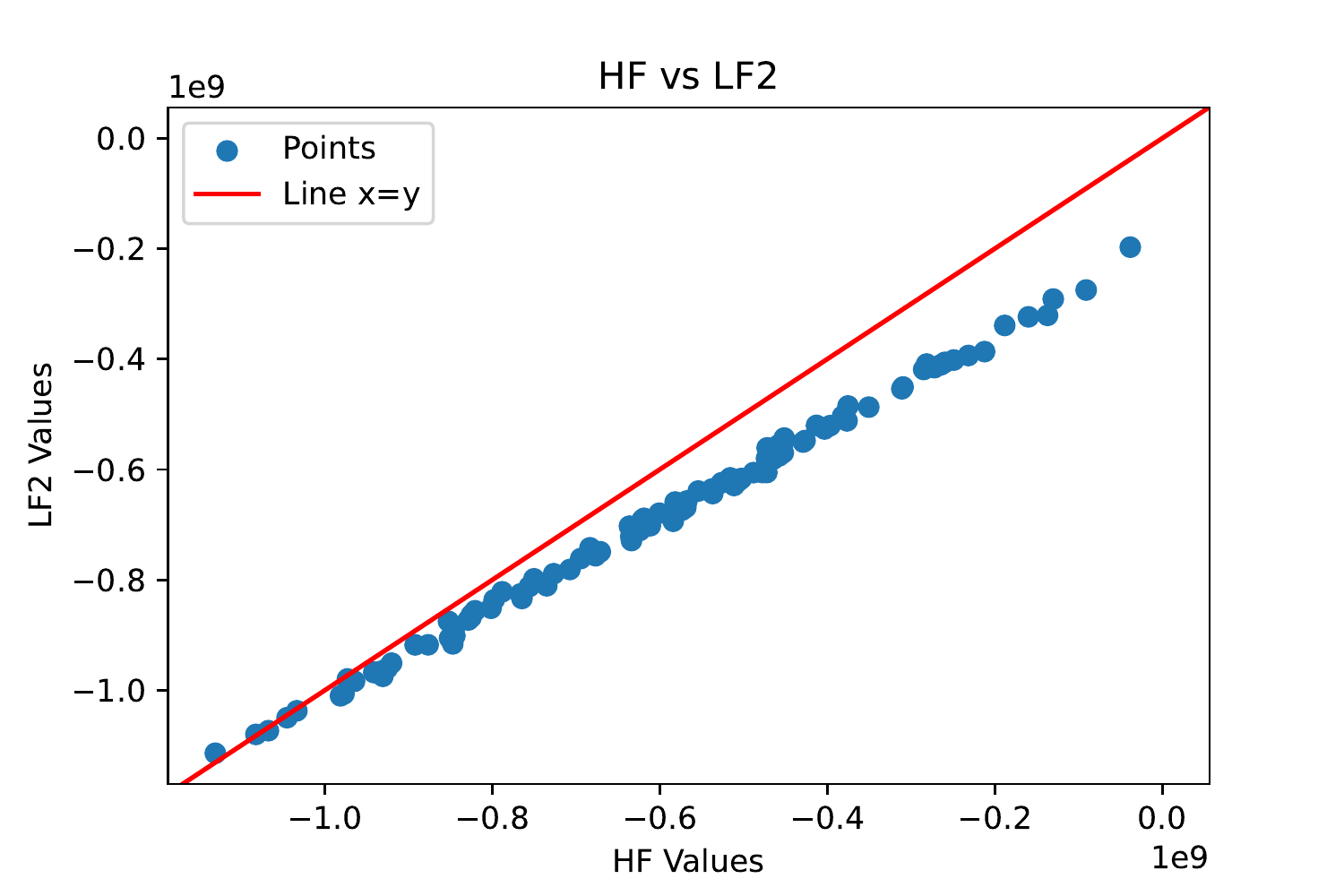} % Reduce the figure size so that it is slightly narrower than the column. Don't use precise values for figure width.This setup will avoid overfull boxes.
\caption{Scatter plot of stresses in the SiC layer computed using 2-D fine mesh (represented on x-axis with ``HF values'') and those computed using 2-D coarse mesh (represented on y-axis with ``LF2 values'').}
\label{fig:triso_stresses}
\end{figure}

The deviations in the failure probabilities reported in Tables \ref{table2} and \ref{table3} needs some discussion. For Table \ref{table2}, the HF model is a 2-D coarse mesh, and for Table \ref{table3}, the HF model is a 2-D fine mesh. The mesh density can result in differences in the failure probability estimates due to the quality of stress estimation in the SiC layer. For example, Figure \ref{fig:triso_stresses} compares the stresses in the SiC layer computed using 2-D fine mesh (represented on x-axis with ``HF values'') with those computed using 2-D coarse mesh (represented on y-axis with ``LF2 values''). It is noticed that the 2-D coarse mesh is underestimating the stress values resulting in an underestimation of the computed failure probability. In any case, what is clear from Tables \ref{table2} and \ref{table3} is that given a HF model and suite of LF models, the multifidelity active learning method correctly estimates the failure probability given its agreement to the reference method.

\section{Summary and conclusions}

The TRISO nuclear fuel is being proposed for use in several advanced reactor concepts and determining its reliability during reactor operation is critical. However, the computational model for TRISO fuel is expensive which makes determining its failure probability a time consuming task. We presented a multifidelity active learning procedure based on Gaussian Processes to accelerate the failure probability estimation given a HF model and suite of LF models. We applied this multifidelity active learning procedure to two cases: (1) one LF and one HF TRISO models; and (2) two LF and one HF TRISO fuel models. In both these cases, the presented active learning procedure accurately estimated the fuel failure probability in comparison to the reference solution while bringing about two orders of magnitude reduction in the computational cost assessed by the number of times the HF model is called.

% \section{Ethical statement.}
% You can write a statement about the potential ethical impact of your work, including its broad societal implications, both positive and negative. If included, such statement must be written in an unnumbered section titled \emph{Ethical statement}.

\section{Acknowledgments}

This manuscript has been authored by Battelle Energy Alliance, LLC under Contract No.~DE-AC07-05ID14517 with the US Department of Energy. The United States Government retains and the publisher, by accepting the article for publication, acknowledges that the United States Government retains a nonexclusive, paid-up, irrevocable, worldwide license to publish or reproduce the published form of this manuscript, or allow others to do so, for United States Government purposes.

\bibliography{main}

% Use \bibliography{yourbibfile} instead or the References section will not appear in your paper
% \nobibliography{aaai23}

\end{document}